\documentclass[twocolumn]{revtex4-2}
\usepackage{amsmath}
\begin{document}

\title{The first-order orbital equation}
\author{Maurizio M. D'Eliseo$^a$}
\affiliation{Osservatorio S. Elmo, Via A. Caccavello 22, 80129 Napoli, Italy}
\date{Received 27 February 2006; accepted 15 December 2006}

\begin{abstract}
We derive the first-order orbital equation employing a complex variable formalism. We then
examine Newton's theorem on precessing orbits and apply it to the perihelion shift of an elliptic
orbit in general relativity. It is found that corrections to the inverse-square gravitational force law
formally similar to that required by general relativity were suggested by Clairaut in the 18th
century.
\end{abstract}

\maketitle

\section{INTRODUCTION}
Almost all classical mechanics textbooks derive the elliptical orbit of the two-body planetary problem by means of well known methods. In this paper we derive the first-order orbital equation by using the complex variable formalism. The latter is a useful tool for studying this old problem from a new perspective. From the orbital equation we can extract all the properties of elliptic orbits. Newton's theorem of revolving orbits,$^1$ which establishes the condition for which a closed orbit revolves around the center of force, has a wide range of applicability, and its application to an inverse-square force allows us to apply the first-order orbital equation.

A revolving precessing ellipse reminds us of the general relativistic perihelion shift of the planet Mercury. We explain why an approximate general relativistic force found by Levi-Civita$^2$ in his lectures on general relativity gives the same perihelion shift of the $r^{-4}$ general relativistic force derived in textbooks. Our results suggest an interesting link with the work of the 18th century scientist Clairaut.$^{3,4}$

We identify the plane of the motion of the gravitational two-body problem with the complex plane.$^5$ An object of mass $M$ is at the origin. The position $x$, $y$ of the second object of mass $m$ is given by $r = x + iy$, and the equation of motion can be expressed as
\begin{align}
\ddot{r} = -\frac{\mu}{r^3}r = -\frac{\mu e^{i\phi}}{r^2},
\end{align}
where $\mu = G(m + M)$, $r = re^{i\phi} = r(\cos\phi + i\sin\phi)$, $r^* = re^{-i\phi}$ is the complex conjugate of $r$, $r = |r| = \sqrt{rr^*} = \sqrt{x^2 + y^2}$, and $\phi=\phi(t)$ is the true longitude, that is, the point $(x,y)$ has the polar form $re^{i\phi}$, where $r$ is the modulus and $\phi$ is its argument. From $r = re^{i\phi}$ we have by differentiation with respect to time
\begin{align}
\dot{r} = \dot{r}e^{i\phi} + ir\dot{\phi}e^{i\phi},
\end{align}
where $\dot{r}^* = \dot{r}^*$. The solution of Eq. (1) requires knowledge of the functions $r(\phi)$ and $\phi(t)$, but we are interested here only in the determination of the function $r(\phi)$, which describes the geometry of the orbit.

We denote by $\text{Re}(r)$ and $\text{Im}(r)$ the real and the imaginary parts of $r$, respectively. Thus $\text{Re}(r) = (r + r^*)/2 = x = r\cos\phi$ and $\text{Im}(r) = (r - r^*)/(2i) = y = r\sin\phi$. It is useful to consider complex variables as vectors starting from the origin so that $\text{Re}(A)$ is the component of $A$ along the $x$ (real) axis, and $\text{Im}(A)$ is the component along the $y$ (imaginary) axis.

\section{THE FIRST-ORDER ORBITAL EQUATION}
If we multiply Eq. (1) by $r^*$, we have
\begin{align}
\ddot{r}r^* = -\frac{\mu rr^*}{r^3} = -\frac{\mu r^2}{r^3} = -\frac{\mu}{r}.
\end{align}
If we take the imaginary part of both sides of Eq. (3), we find
\begin{align}
\text{Im}(\ddot{r}r^*) = -\text{Im}\left(\frac{\mu}{r}\right) = 0.
\end{align}
It is easy to verify that
\begin{align}
\frac{d}{dt}\text{Im}(\dot{r}r^*) = \text{Im}(\ddot{r}r^*) + \text{Im}(\dot{r}\dot{r}^*).
\end{align}
We have $\text{Im}(\dot{r}\dot{r}^*) = 0$ because the term in brackets is the square of the module $|\dot{r}|$, a real quantity. Then from Eqs. (5) and (4) we write
\begin{align}
\frac{d}{dt}\text{Im}(\dot{r}r^*) = 0.
\end{align}
Equation (6) implies that $\text{Im}(\dot{r}r^*)$ is time independent. We denote its real value by $\ell$ and write
\begin{align}
\text{Im}(\dot{r}r^*) = \frac{\dot{r}r^* - \dot{r}^*r}{2i} = \ell.
\end{align}
Equation (7) is the area integral.

This derivation holds for any central force $f(r)e^{i\phi} = f(r)r/r$. We substitute Eq. (2) into Eq. (7) and find the fundamental relation $\ell = r^2\dot{\phi}$, which can be cast in three equivalent forms:
\begin{align}
\frac{1}{r^2} &= \frac{\dot{\phi}}{\ell}, \\
dt &= \frac{r^2}{\ell}d\phi, \\
\frac{d}{dt} &= \frac{\ell}{r^2}\frac{d}{d\phi}.
\end{align}
We can rewrite Eq. (1) using Eq. (8a) as
\begin{align}
\ddot{r} = -\frac{\mu}{\ell}\dot{\phi}e^{i\phi} = -\frac{i\mu}{\ell}\frac{d}{dt}e^{i\phi},
\end{align}
or
\begin{align}
\frac{d}{dt}\left[\dot{r} - \frac{i\mu}{\ell}e^{i\phi}\right] = 0.
\end{align}
The expression in parentheses is complex and constant in time. For convenience we denote it as $i\mu e/\ell$. The reason for this choice will soon be apparent. We have thus deduced the Laplace integral$^6$
\begin{align}
\dot{r} = \frac{i\mu}{\ell}e^{i\phi} + e\mu,
\end{align}
where $\dot{r}$ is the orbital velocity, $e = e\exp(i\varpi)$ is a complex constant that we will call the eccentricity vector, and $e$ is the scalar eccentricity. The vector $e$ is directed toward the perihelion, the point on the orbit of nearest approach to the center of force, and $\varpi$ is the argument of the perihelion.

If we use the area integral to eliminate the explicit presence of $t$ in Eq. (11), we obtain a relation between $r$ and $\phi$. One way to integrate Eq. (1) twice with respect to time is to substitute into Eq. (7) the expression for $\dot{r}$ given by Eq. (11):
\begin{align}
\ell &= \text{Im}(\dot{r}r^*) = \text{Im}\left(\frac{i\mu}{\ell}r + e\mu r^*\right) \\
&= \frac{\mu}{\ell}r + \text{Im}(ie\mu r^*) \\
&= \frac{\mu}{\ell}r(1 + \text{Re}(ee^{-i\phi})),
\end{align}
from which we can solve for $r$
\begin{align}
r = \frac{\ell^2/\mu}{1 + e\cos(\phi - \varpi)}.
\end{align}
Equation (13) is the relation in polar coordinates of the orbit, which is a conic section of eccentricity $e$ with a focus at the origin. If the orbit is an ellipse $(0 \leq e < 1)$, we have the relation $\ell^2/\mu = a(1 - e^2)$, where $a$ is the semi-major axis which lies on the apse line.$^7$ Thus the names given earlier to $e = |e|$ and $\varpi$ are justified.

Another way to find the orbit is to transform the time derivative into a $\phi$ derivative. We start from Eq. (11), which is a function of $\phi$ instead of time. From Eq. (8c) we obtain
\begin{align}
\dot{r} = \frac{\ell}{r^2}r' = \frac{i\mu}{\ell}e^{i\phi} + e\mu,
\end{align}
where a prime denotes differentiation with respect to $\phi$. Then
\begin{align}
r' = r'e^{i\phi} = r' + ire^{i\phi} = \frac{i\mu r^2}{\ell^2}e^{i\phi} + e\mu r^2.
\end{align}
If we multiply by $e^{-i\phi}$, we obtain the complex Bernoulli equation$^8$
\begin{align}
r' + ir = \frac{i\mu r^2}{\ell^2}(1 + ee^{-i\phi}).
\end{align}
If we take the imaginary and the real parts of both sides of Eq. (16), we obtain the orbit $r(\phi)$ and its derivative $r'(\phi)$, respectively. However, it is better to change the dependent variable, so we divide both sides by $r^2$, make the variable change $r \rightarrow 1/u$, and multiply by $i$. We find
\begin{align}
u' + iu = \frac{\mu}{\ell^2}(1 + ee^{-i\phi}).
\end{align}
Despite its heterogeneous nature, it is convenient to write the left-hand side of Eq. (17) in terms of $u = u(\phi) = u' + iu$, so that we have
\begin{align}
\mathcal{D}u = \frac{\mu}{\ell^2}(1 + ee^{-i\phi}),
\end{align}
which we call the first-order orbital equation.

From Eq. (18) we can immediately deduce the orbit and its apsidal points. The orbit is given by the real part of $u$,
\begin{align}
\text{Re}(u) = \frac{1}{r} = \frac{\mu}{\ell^2}(1 + \text{Re}(ee^{-i\phi})) = \frac{\mu}{\ell^2}(1 + e\cos(\phi - \varpi)).
\end{align}
The apsidal points are determined from the condition $\text{Im}(u) = 0$ because $r'(\phi) = 0$ at these points. If $\text{Im}(u) = 0$ for every value of $\phi$, then $e = 0$, and we have a circular orbit with $u = \text{Re}(u) = \mu/\ell^2$. If $0 < e \leq 1$, then Eq. (18) gives the position of the two apsidal points $r_{\text{min}}$ and $r_{\text{max}}$. At these points we have
\begin{align}
\text{Im}(u) = -\frac{\mu}{\ell^2}e\sin(\phi - \varpi) = 0,
\end{align}
so that, by considering the derivative $\text{Im}(u)'(\phi) = u'(\phi)$, we find $r_{\text{min}}$ when $\phi + 2\pi n = \varpi$ and $r_{\text{max}}$ when $\phi + \pi(2n + 1) = \varpi$, where $n = 0, 1, 2, \ldots$.

If we denote by $D$ the differential operator $d/d\phi$, Eq. (18) may be written in operator form as
\begin{align}
\mathcal{D}u = (1 + iD)u = \frac{\mu}{\ell^2}(1 + ee^{-i\phi}).
\end{align}
By multiplying both sides on the left by $(1 - iD)$, we obtain
\begin{align}
(1 - iD)(1 + iD)u = (D^2 + 1)u = u'' + u = \frac{\mu}{\ell^2},
\end{align}
which is Binet's orbit equation.$^9$

\section{THE PRECESSING ELLIPSE}
The two-body solution we have found together with the appropriate corrections due to the presence of other bodies does not account for the observed residual precession of the planetary perihelia.$^{10}$ An explanation in classical terms is that a small additional force acts on all the planets causing precession.

All perturbing central forces of the type $F \propto r^{-m}e^{i\phi}$, with $m \neq 3$, produce a secular motion of the apse of an elliptical orbit. Conversely, from the observed planetary apse motion we can deduce by Newton's theorem the presence of a perturbing inverse-cube force $F \propto r^{-3}e^{i\phi}$. This result was obtained by Newton in more general terms using the following reasoning.$^1$

Consider a closed orbit determined by the centripetal force $-f(r)e^{i\phi}$. If we let $\tilde{r} = r$, where $\lambda \neq 1$ is an arbitrary real constant, we will obtain the same orbit as for $\lambda = 1$, but revolving around the center of force (the two orbits are coincident when $\phi = 0$). From the area integral of the first orbit, $r^2\dot{\phi} = \ell$, we obtain $\tilde{r}^2\dot{\tilde{\phi}} = \tilde{\ell}$, which we write as $\tilde{r}^2\dot{\tilde{\phi}} = \tilde{\ell}$. This integral is for the centripetal force $-\tilde{f}(\tilde{r})e^{i\phi}$. The radial equations of these two orbits with the same $r(t)$ are
\begin{align}
\ddot{r} - r\dot{\phi}^2 &= -f(r), \\
\ddot{r} - r\dot{\tilde{\phi}}^2 &= -\tilde{f}(r),
\end{align}
from which we obtain
\begin{align}
\tilde{f}(r) - f(r) &= r\dot{\tilde{\phi}}^2 - \dot{\phi}^2 = r\left(\frac{\tilde{\ell}^2}{r^4} - \frac{\ell^2}{r^4}\right) \\
&= \frac{\ell^2(\lambda^2 - 1)}{r^3}.
\end{align}
The extra radial force is outward or inward depending on whether $\lambda$ is greater or less than unity.

Thus far the force $f(r)$ is arbitrary, but if we specialize to the inverse-square gravitational force, then the first-order orbital equation (18) with the perturbing inverse-cube force
\begin{align}
\frac{\ell^2(\lambda^2 - 1)}{r^3}e^{i\phi} = \ell^2(\lambda^2 - 1)u^3e^{i\phi}
\end{align}
takes the form
\begin{align}
\mathcal{D}u = \frac{\mu}{\ell^2\lambda}(1 + ee^{-i\phi}).
\end{align}
Hence
\begin{align}
u = \text{Re}(u) = \frac{\mu}{\ell^2\lambda} + \frac{\mu}{\ell^2\lambda}e\cos(\phi - \varpi)
\end{align}
is an ellipse precessing around the focus with an angular velocity proportional to the radius vector. This description becomes more accurate as $\lambda$ approaches unity. The apsidal points are given by $\text{Im}(u) = 0$. From Eq. (26) and $e = e\exp(i\varpi)$, we have at the apsidal points
\begin{align}
\sin(\lambda\phi - \varpi) = 0.
\end{align}
In particular we have $r_{\text{min}}$ when
\begin{align}
\lambda\phi - \varpi = -\pi \Rightarrow \phi + 1 - \lambda\phi = 0,
\end{align}
so that after one complete revolution the angular perihelion shift is
\begin{align}
\Delta\phi = 2\pi(1 - \lambda).
\end{align}
If $\lambda > 1$, then $\Delta\phi < 0$ and the shift is positive, while for $\lambda < 1$ we have $\Delta\phi > 0$ and the shift is negative.

\section{GENERAL RELATIVITY}
The foregoing considerations have a direct application in general relativity. The general relativistic Binet's orbit equation, which is obtained from the geodesic equation in the Schwarzschild space-time, is$^{11}$
\begin{align}
u'' + u = \frac{\mu}{\ell^2} + 3\rho_g u^2,
\end{align}
where $\rho_g = GM/c^2 \ll \ell^2/\mu$ is the gravitational radius of the central body, and $c$ is the speed of light. The corresponding equation of motion is
\begin{align}
\ddot{r} = -\frac{\mu}{r^2}e^{i\phi} - \frac{3\rho_g\ell^2}{r^4}e^{i\phi}
\end{align}
and we see that general relativity introduces an effective perturbative $r^{-4}$ force.

From Eq. (31) or Eqs. (11) and (32) we can deduce$^{12}$ the standard formula for the perihelion shift given in general relativity by
\begin{align}
\Delta\phi = \frac{6\pi\rho_g}{\ell^2/\mu} = \frac{6\pi\rho_g}{a(1 - e^2)}.
\end{align}
If we equate Eqs. (30) and (33) and solve for $\lambda$, we obtain $\lambda \approx 1 - 6\pi\rho_g/\ell^2$, and by using Eq. (25) we obtain the inverse-cube perturbation that gives the same perihelion shift as predicted by general relativity:
\begin{align}
F = -\frac{\mu 6\pi\rho_g}{r^3}e^{i\phi}.
\end{align}
That is, we can obtain the same precession using either an $r^{-3}$ or $r^{-4}$ perturbative force.

Levi-Civita obtained an effective $r^{-3}$ force in general relativity using a method based on a new form of Hamilton's principle$^2$ devised to go smoothly from the classical equation of motion to the Einstein field equation. His approximation is not as general as the usual $r^{-4}$ effective force because it does not produce the bending of light rays, a subject that Levi-Civita treated with another ingenious approximation.

Binet's equation for the $r^{-3}$ perturbative force, obtained from the equation of the motion by the use of Eq. (8c) and the variable change $r \rightarrow 1/u$, is
\begin{align}
u'' + \left(1 - \frac{6\pi\rho_g}{\ell^2}\right)u = \frac{\mu}{\ell^2}.
\end{align}
The null-geodesic equation of light rays requires that we formally put $\mu/\ell^2 = 0$ in Eq. (35),$^{13}$ so that it becomes
\begin{align}
u'' + u = 0.
\end{align}
The solution of Eq. (36) is $k\sin\phi$ where $k = \text{const}$ and $\phi \neq 0$. In terms of the radius $r = 1/u$, the solution becomes $r\sin\phi = 1/k$. Because $r\sin\phi$ is the Cartesian coordinate $y$, the solution represents a straight line parallel to the $x$ axis, so that the light ray is not deflected at all by the sun's gravitational field in this approximation.

As we have seen, the weak-field approximation of general relativity adds an effective $r^{-3}$ force or a $r^{-4}$ force depending on the approximation method used to Newton's inverse square force to explain the perihelion motion. It is interesting that these results were proposed in the mid-18th century by the mathematician and astronomer Clairaut who proposed the addition of a small $r^{-n}$ force to the $r^{-2}$ gravitational force to explain the swift motion of the lunar perigee. In particular, he examined the influence of both $r^{-4}$ and $r^{-3}$ terms.$^{14}$ It was later recognized by Clairaut that this addition was unnecessary, because a purely $r^{-2}$ force law could completely explain the motion of the Moon. The apparently anomalous secular motion of the perigee was due to discarded noncentral force terms in the process of successive approximations.$^{15}$ No doubt Clairaut would have again made his suggestion if he had known about the anomalous motion of Mercury's perihelion. Without a new first principles gravitational theory$^{16}$ he probably would have employed a phenomenological approach and introduced one of the two forces by empirically adjusting the numerical factors.

\begin{footnotesize}
$^{a}$Electronic mail: s.elmo@mail.com

$^1$S. Chandrasekhar, \textit{Newton's Principia for the Common Reader} (Clarendon, Oxford, 1995), pp. 184--187. In particular, Proposition XLIV-Theorem XIV: The difference of the forces, by which two bodies may be made to move equally, one in a fixed, the other in the same orbit revolving, varies inversely as the cube of their common altitudes.

$^2$T. Levi-Civita, \textit{Fondamenti di Meccanica Relativistica} (Zanichelli, Bologna, 1929), p. 123.

$^3$A. C. Clairaut, ``Du Systeme du Monde, dan les principes de la gravitation universelle,'' Histoires de l'Academie Royale des Sciences, mem. 1745 and Ref. 4.

$^4$We have reproduced many papers of historical interest at gallica.bnf.fr/.

$^5$T. Needham, \textit{Visual Complex Analysis} (Oxford U. P., New York, 1999).

$^6$The Laplace integral can be found in P. S. Laplace, \textit{Ouvres} (Gauthier-Villars, Paris, 1878), Tome 1, p. 181, formula P. To obtain Eq. (11), we need to use $z = 0$, $c = xy\dot{}y - \dot{x}y = \ell$, $f = \mu \text{Re}(e)$, and $f' = \mu\text{Im}(e)$, and add the first relation to the second one multiplied by $-i$ (see Ref. 4). To keep the customary notation we use the same letter $e$ for the eccentricity and for the complex exponential.

$^7$A parallel treatment of the two-body problem with vectorial methods is given by V. R. Bond and M. C. Allman, \textit{Modern Astrodynamics} (Princeton U. P., Princeton, NJ, 1998).

$^8$R. E. Williamson, \textit{Introduction to Differential Equations} (McGraw-Hill, New York, 1997), p. 84.

$^9$R. d'Inverno, \textit{Introducing Einstein's Relativity} (Oxford U. P., New York, 2001), p. 194.

$^{10}$For a complete calculation of all the perturbing effects see M. G. Stewart, ``Precession of the perihelion of Mercury's orbit,'' Am. J. Phys. \textbf{73}, 730--734 (2005).

$^{11}$Reference 9, p. 196.

$^{12}$See, for example, B. Davies, ``Elementary theory of perihelion precession,'' Am. J. Phys. \textbf{51}, 909--911 (1983); N. Gauthier, ``Periastron precession in general relativity,'' ibid. \textbf{55}, 85--86 (1987); T. Garavaglia, ``The Runge-Lenz vector and Einstein perihelion precession,'' ibid. \textbf{55}, 164--165 (1987); C. Farina and M. Machado, ``The Rutherford cross section and the perihelion shift of Mercury with the Runge-Lenz vector,'' ibid. \textbf{55}, 921--923 (1987); D. Stump, ``Precession of the perihelion of Mercury,'' ibid. \textbf{56}, 1097--1098 (1988); K. T. McDonald, ``Right and wrong use of the Lenz vector for non-Newtonian potentials,'' ibid. \textbf{58}, 540--542 (1990); S. Cornbleet, ``Elementary derivation of the advance of the perihelion of a planetary orbit,'' ibid. \textbf{61}, 650--651 (1993); B. Dean, ``Phase-plane analysis of perihelion precession and Schwarzschild orbital dynamics,'' ibid. \textbf{67}, 78--86 (1999).

$^{13}$R. Adler, M. Bazin, and M. Shiffer, \textit{Introduction to General Relativity}, 2nd ed. (McGraw-Hill, New York, 1975), p. 216, Eq. 6.149.

$^{14}$A. C. Clairaut, ``Du Systeme du Monde, dan les principes de la gravitation universelle,'' Histoires de l'Academie Royale des Sciences, mem. 1745, p. 337: Clairaut wrote that ``The moon without doubt expresses some other law of attraction than the inverse square of the distance, but the principal planets do not require any other law. It is therefore easy to respond to this difficulty, and noting that there are an infinite number of laws which give an attraction which differs very sensibly from the law of the squares for small distances, and which deviates so little for the large, that one cannot perceive it by observations. One might regard, for example, the analytic quantity of the distance composed of two terms, one having the square of the distance as its divisor, and the other having the square square.'' On p. 362, Clairaut examined the effect of a perturbing inverse-cube force. This memoir is dated 15 November 1747 and can be found in Ref. 4. Clairaut was also the first to introduce a revolving ellipse as a first approximation to the motion of the moon. This idea is sometimes called Clairaut's device or Clairaut's trick.

$^{15}$See F. Tisserand, \textit{Traité de Mecanique Celeste III} (Gauthier-Villars, Paris, 1894), p. 57; reproduced at Ref. 4.

$^{16}$Clairaut was also a first-class geometer, specializing in curvature. See his \textit{Recherches sur le courbes a double courbure} at Ref. 4.
\end{footnotesize}

\end{document}